\documentclass[aps,prl,10pt,twocolumn,superscriptaddress,nobibnotes]{revtex4-2}

\usepackage[charter]{mathdesign}
\DeclareSymbolFont{usualmathcal}{OMS}{cmsy}{m}{n}
\DeclareSymbolFontAlphabet{\mathcal}{usualmathcal}
\usepackage{amsmath}
\usepackage[pdftex]{graphicx}
\usepackage{microtype}
\usepackage{dsfont}
\usepackage{bm}
\usepackage[svgnames]{xcolor}
\usepackage[colorlinks=True,
            linkcolor=DarkRed,
            citecolor=VioletRed,
            urlcolor=MediumBlue,
	        pdfstartview=FitH,
            bookmarks=False,pdfpagemode=UseNone]{hyperref}
\usepackage[left=1.85cm,right=1.85cm,top=2cm,bottom=2cm]{geometry}
\usepackage{lipsum}
\usepackage{titlesec}
\titleformat{\section}
    {\normalfont\bfseries}
    {\thesection}{1em}{}
\titlespacing*{\section}{0pt}{3.5ex plus 1ex minus .2ex}{0ex}
\titleformat{\subsection}
    {\normalfont\bfseries}
    {\thesubsection}{1em}{}
\titlespacing*{\subsection}{0pt}{3.25ex plus 1ex minus .2ex}{0ex}
\titleformat{\subsubsection}
    {\normalfont\itshape}
    {\thesubsubsection}{1em}{}
\titlespacing*{\subsubsection}{0pt}{3.25ex plus 1ex minus .2ex}{0ex}

\counterwithout{equation}{section}

\newcommand{\vect}[1]{\bm{#1}}
\newcommand{\panel}[1]{{\fontfamily{phv}\selectfont\textbf{#1}}}
\newcommand{\comment}[1]{\relax}

\def\grad{\nabla}

\begin{document}
\title{End state of the experimental black-hole bomb}%

\author{Patrik \v{S}van\v{c}ara}
\email{patrik.svancara@neel.cnrs.fr}
\affiliation{School of Mathematical Sciences, University of Nottingham, NG7 2RD Nottingham, UK}
\affiliation{Nottingham Centre of Gravity, University of Nottingham,
NG7 2RD Nottingham, UK}
\affiliation{Univ. Grenoble Alpes, CNRS, Grenoble INP, Institut N\'{e}el, 38000 Grenoble, France}

\author{Leonardo Solidoro}
\affiliation{School of Mathematical Sciences, University of Nottingham, NG7 2RD Nottingham, UK}
\affiliation{Nottingham Centre of Gravity, University of Nottingham,
NG7 2RD Nottingham, UK}

\author{Pietro Smaniotto}
\affiliation{School of Mathematical Sciences, University of Nottingham, NG7 2RD Nottingham, UK}
\affiliation{Nottingham Centre of Gravity, University of Nottingham,
NG7 2RD Nottingham, UK}

\author{Sam~Patrick}
\affiliation{Department of Physics and Astronomy, University of Manchester, M13 9PL Manchester, UK}
\affiliation{Photon Science Institute, University of Manchester, M13 9PY Manchester, UK}
\affiliation{Department of Physics, King’s College London, University of London, Strand, WC2R 2LS London, UK}

\author{Silvia Schiattarella}
\affiliation{School of Physics and Astronomy, University of Nottingham, NG7 2RD Nottingham, UK}

\author{Maciej T. Jarema}
\affiliation{School of Mathematical Sciences, University of Nottingham, NG7 2RD Nottingham, UK}
\affiliation{Nottingham Centre of Gravity, University of Nottingham, NG7 2RD Nottingham, UK}

\author{Sean M. D. Gregory}
\affiliation{School of Mathematical Sciences, University of Nottingham, NG7 2RD Nottingham, UK}

\author{Vitor S. Barroso}
\affiliation{School of Mathematical Sciences, University of Nottingham, NG7 2RD Nottingham, UK}

\author{Maur\'icio Richartz}
\affiliation{Centro de Matem\'atica, Computa\c c\~ao e Cogni\c c\~ao,
Universidade Federal do ABC (UFABC), 09210-170 Santo Andr\'e, S\~ao Paulo, Brazil}

\author{Anastasios Avgoustidis}
\affiliation{School of Physics and Astronomy, University of Nottingham, NG7 2RD Nottingham, UK}

\author{Carlo F. Barenghi}
\affiliation{School of Mathematics, Statistics and Physics, Newcastle University, NE1 7RU Newcastle upon Tyne, UK}

\author{Ruth Gregory}
\affiliation{Department of Physics, King’s College London, University of London, Strand, WC2R 2LS London, UK}
\affiliation{Perimeter Institute, 31 Caroline Street North, N2L 2Y5 Waterloo, Ontario, Canada}

\author{Silke Weinfurtner}
\affiliation{School of Mathematical Sciences, University of Nottingham, NG7 2RD Nottingham, UK}
\affiliation{Nottingham Centre of Gravity, University of Nottingham, NG7 2RD Nottingham, UK}
\affiliation{Department of Physics and Astronomy, University of Manchester, M13 9PL Manchester, UK}
\affiliation{Photon Science Institute, University of Manchester, M13 9PY Manchester, UK}


\begin{abstract}
\vspace{1.5em} 
\noindent
Rotating black holes can amplify incident waves through superradiant scattering. When these waves are confined, repeated amplification gives rise to the black-hole bomb instability, whose nonlinear evolution remains poorly understood despite its central role in models of bosonic clouds around astrophysical black holes. Here, we reproduce the black-hole bomb mechanism in a laboratory setting using a gravity simulator based on a draining vortex in superfluid helium. Surface waves propagating on the superfluid interface experience an effective rotating spacetime and undergo repeated superradiant amplification within a cylindrical cavity. By tuning the temperature and flow parameters, we achieve exponential growth of a low-frequency resonant mode, followed by the arrest of the instability and the formation of a long-lived non-equilibrium steady state. Using spatially and temporally resolved measurements, we identify nonlinear frequency shifts, harmonic generation, and coherent three- and four-wave mixing that redistribute energy among interacting modes. This novel end state of the experimental black-hole bomb highlights the role of nonlinear wave interactions in quenching the runaway growth expected from linear theory and governing the system's late-time dynamics. Our results establish a laboratory framework for investigating the nonlinear evolution of black-hole bombs, with implications for analogous phenomena involving ultralight bosonic fields around rotating black holes.
\end{abstract}

\maketitle

The spacetime structure associated with a rotating black hole permits the extraction of its rotational energy through the Penrose process \cite{penrose1971}. In this mechanism, an object approaching the black hole splits into two fragments: while one fragment is captured by the black hole, the other escapes with energy exceeding that of the original, whole object. The same amplification occurs for continuous fields (waves) scattering off a rotating black hole. This principle, known as rotational superradiance \cite{zel1971}, underpins the black-hole bomb mechanism \cite{press1972}, wherein a field confined between the black hole and a reflective boundary undergoes repeated, runaway amplification.

A closely related phenomenology arises in recent studies of environmental effects around black holes. Partial confinement of superradiant modes may occur in the presence of surrounding plasma and dark photons~\cite{cannizzaro2021a,cannizzaro2021b}, axion fields~\cite{kiczek2021}, or scalar hair~\cite{lei2023}. In these settings, the field mass or a medium-induced dispersion relation act as effective confining mechanisms, enabling long-lived bosonic clouds to grow by extracting energy and angular momentum from rapidly rotating black holes \cite{Brito:2015oca}.

\begin{figure*}[htbp]
    \centering
    \includegraphics[width=\textwidth]{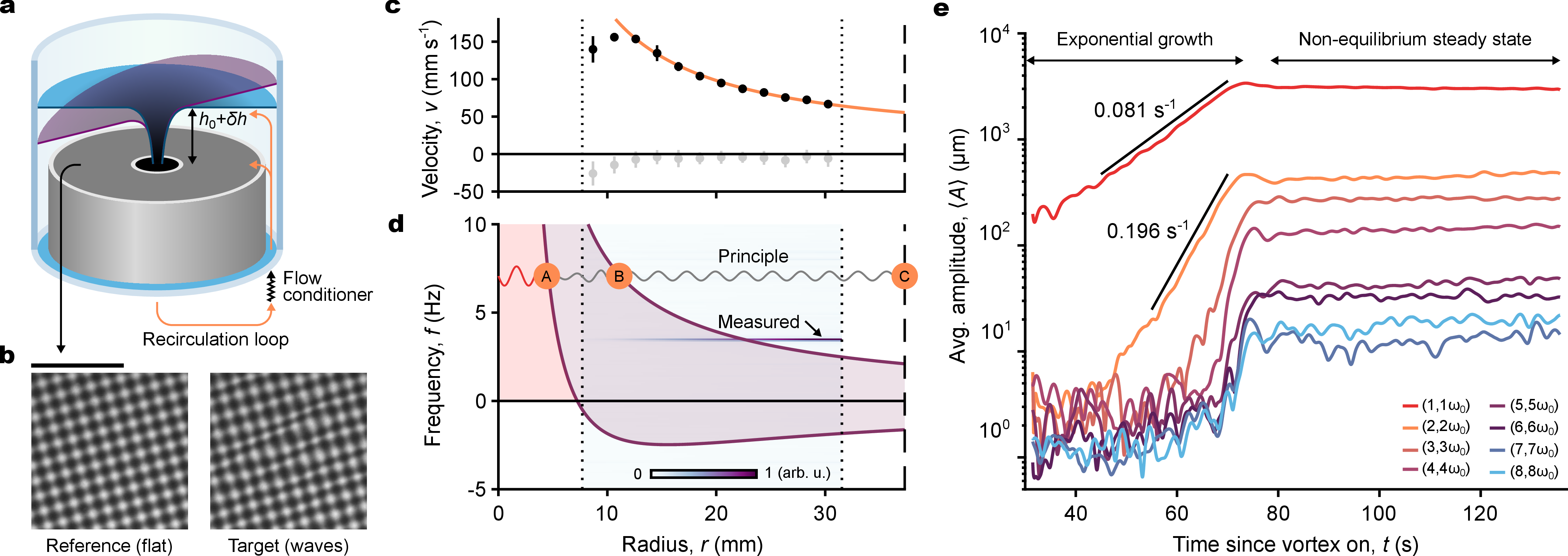}
     \caption{{\bf Experimental realisation of the black-hole bomb.} %
     \panel{a} A draining vortex forms in superfluid helium-4 within the cylindrical experimental zone above a $10$-mm-diameter opening. A spinning propeller underneath ensures steady recirculation of the superfluid (yellow arrows), while a flow conditioner (black zig-zag arrow) minimises solid-body rotation \cite{svancara2024}. Initially, a hollow-core vortex appears (blue-shaded surface), which eventually develops centimetre-scale sloshing (violet surface). %
    \panel{b} The free surface of superfluid helium is imaged against a regular pattern (Reference, left), which covers the dark grey shaded area in panel a. Interface fluctuations $\delta h$ are reconstructed by comparing the reference with wave-induced pattern deformations (Target, right). Scale bar, 5~mm. %
    \panel{c}~Azimuthal (black) and radial (grey) velocity components of the vortex flow as a function of radius, averaged over $2$-mm intervals. The orange line fits Eq.~\eqref{eq:velocity} and error bars indicate one standard deviation. The accessible field of view is marked by black dotted lines, and the outer boundary by a black dashed line. %
    \panel{d} The effective scattering potential for azimuthal number $m = 1$ (violet lines) delimits the region admitting only evanescent waves (violet-shaded area). The sinusoidal line (Principle) illustrates tunnelling across the potential barrier that fuels the black-hole bomb. The background map shows the radially resolved amplitude spectrum of surface waves, dominated by the sloshing mode (Measured), oscillating at $\omega_0/(2\pi) = 3.47$~Hz.  %
    \panel{e} Frequency- and $m$-filtered, radially-averaged amplitudes $\left\langle A \right\rangle$ for the sloshing mode $(m=1,\omega=\omega_0)$ and its multiples (see legend) reveal a phase of exponential growth followed by a non-equilibrium steady state. Solid black lines indicate the best linear fits in log-linear scale, vertically shifted for clarity, with the growth rates shown alongside.%
   }
    \label{fig:experiment}
\end{figure*}

A complementary approach exploits analogue gravity systems, where black-hole bomb dynamics can be investigated under controlled laboratory conditions \cite{barcelo2001}. In these systems, waves propagating on an inhomogeneous medium experience an effective spacetime geometry that can reproduce horizons, ergoregions, and other black-hole phenomena. Superradiant amplification, required for the instability to occur, has been observed across several experimental platforms: surface waves around a draining water vortex \cite{torres2017}, light in nonlinear media \cite{braidotti2022}, and electromagnetic waves in an electric circuit \cite{braidotti2024}. By incorporating an LC resonator into the latter system, repeated reflection of the superradiant wave was achieved, resulting in the first experimental realisation of the black-hole bomb \cite{cromb2025}. In this setup, the runaway amplification was finite in duration, as the energy of the rotating object (a rotor in a modified three-phase electric motor) was depleted below the superradiant threshold.

We realise a black-hole bomb in a rotating analogue spacetime generated by a draining vortex in superfluid helium-4 \cite{svancara2024}. Unique hydrodynamic properties of this quantum liquid enable us to establish a sufficiently powerful vortex to support the repeated superradiant amplification of superfluid surface waves, which act as the trapped field \cite{patrick2024primer}. Unlike previous reports, we achieve the black-hole bomb's end state, where nonlinear effects halt the seemingly runaway amplification and drive the system into a long-lived non-equilibrium steady state. In what follows, we identify nonlinear frequency shifts and uncover specific nonlinear wave interactions quenching the instability and redistributing energy among resonant modes.

\section*{Realising the black-hole bomb}

The experimental set-up, schematically shown in Fig.~\ref{fig:experiment}a, is partially submerged in a bath of superfluid helium, such that its depth in the experimental zone reaches $h_0 = 21.6$~mm. The zone is enclosed by a cylindrical container, which acts as an impenetrable boundary for surface waves at constant radius $r_B = 37.3~\mathrm{mm}$. The draining vortex forms once a steady recirculation loop of superfluid helium is established within the set-up (yellow arrows; see Methods). This vortex (blue-shaded surface) possesses a thin, funnel-shaped core that extends between the liquid's free surface and the 10-mm-diameter drain at the bottom of the experimental zone. Crucially, a rotary sloshing wave precessing around the vortex core soon develops on the free surface, reaching a centimetre-scale amplitude (violet-shaded surface).
Our analysis shows that this excitation is the outcome of the black-hole bomb instability. Its realisation is made possible due to superradiant amplification of the wave near the vortex, the wave reflection at the outer boundary, and the relatively low damping of surface waves. The latter is due to the extremely small kinematic viscosity $\nu$ of superfluid helium. At 1.70~K, the temperature at which our experiments took place, $\nu = 8.9\times 10^{-9}~\mathrm{m^2\,s^{-1}}$ \cite{donnelly1998} and the characteristic dissipation time is on the order of $1/(\nu k^2) \approx 100$~s for wavenumber $k = 1~\mathrm{mm^{-1}}$, which is a timescale comparable with the duration of the experiment.

Although the sloshing wave is clearly visible through the walls of the employed glass cryostat, we implement a quantitative detection of surface waves through a flat viewport located above the experimental zone. We specifically image the superfluid interface against a periodic pattern (Fig.~\ref{fig:experiment}b) and capture how the waves distort it, allowing us to digitally reconstruct the surface height fluctuation field $\delta h$ via synthetic Schlieren imaging \cite{moisy2009,wildeman2018}. These fluctuations are captured in an annular field of view excluding the drain and near-boundary regions (Fig.~\ref{fig:experiment}a, grey area). Time- and space-resolved data permit to discern individual oscillating modes in Fourier space: we transform time and the azimuthal coordinate (see Methods), and label these modes by their frequency $f$ and azimuthal number $m$. The latter encodes the periodicity of a given wave around the vortex, with the associated wavelength depending on radius $r$ and being equal to $2\pi r/m$.

Initially, surface waves are driven by ambient mechanical noise, generated by the spinning propeller responsible for the recirculation loop. We already noted \cite{svancara2024} that the characteristic shape of noise-driven frequency spectra can be exploited to extract the background velocity field. Its two components, reflecting the flow in the fluid layer probed by surface waves, are presented in Fig.~\ref{fig:experiment}c. The radial component (grey points) is relatively small and negative, indicating a weak draining of superfluid helium (this effect is likely more substantial deeper below the surface \cite{andersen2003}). In contrast, the azimuthal component (black points) dominates the overall flow field. These data can be fitted (orange line) with the profile,
\begin{equation}
    v_\theta (r) = \frac{C}{r} + \Omega r\,,\label{eq:velocity}
\end{equation}
where $C$ is the circulation of the central vortex and $\Omega$ indicates the angular velocity of the global, solid-body rotation of superfluid helium inside the experimental zone. We intentionally suppress the second, rotational term by implementing a custom flow conditioner within the recirculation loop. As a result, this term contributes less than 6\% to $v_\theta$ in the investigated domain, allowing us to consider an irrotational (vorticity-free) flow field with sub-dominant rotational corrections (see Methods). Long-wavelength surface waves perceive this velocity field, $\vect{v} = v_\theta \,\hat{\vect{\theta}}$, as an effective, (2+1)-dimensional spacetime \cite{visser2005}, whose metric tensor reads
\begin{equation}
      g_{ij} \propto \begin{pmatrix}-c^2+||\bm{v}||^2 & -\bm{v} \\ -\bm{v} & \mathds{1}_{2\times 2}\end{pmatrix},
      \label{eq:metric}
\end{equation}
and $c$ represents the wave propagation speed. Although this formulation does not account for wave dispersion, previous studies \cite{svancara2024,patrick2020} have shown that the generic curved-spacetime phenomenology, including superradiant instabilities discussed here \cite{patrick2024primer}, persist in regimes where dispersion becomes important. 

The first insights into the dynamics of surface waves on a flowing background are offered by the Wentzel–Kramers–Brillouin (WKB) approximation \cite{buhler2014}, treating interface fluctuations as a superposition of plane waves characterised by a spatially varying wave vector $\vect{k}$ and a fixed angular frequency $\omega = 2\pi f$. Without loss of generality, we consider $m > 0$ and admit positive and negative frequencies: waves with $\omega > 0$ co-rotate with the vortex, while those with $\omega < 0$ counter-rotate. By solving the associated dispersion relation (formulated in Methods), we find that propagating, i.e. real-valued, solutions for $\vect{k}$ are not guaranteed throughout the entire experimental domain (see Methods). Instead, we identify a region in the frequency-radius space where only evanescent, i.e. complex-valued, solutions exist. In Fig.~\ref{fig:experiment}d, this region is shown for $m=1$ (i.e., the azimuthal number corresponding to the sloshing mode) as a violet area. Within the WKB framework, radially ingoing and outgoing waves scatter at the edges of the evanescent region, allowing us to interpret it as an effective potential barrier that permits amplitude tunnelling \cite{patrick2024primer,smaniotto2026}.

Due to the background flow, the lower branch of the effective potential (lower violet line) curves upward. Its intersection with the zero-frequency line defines an analogue ergoregion: waves inside the ergoregion are dragged by the flow such that excitations in the pale red area in Fig.~\ref{fig:experiment}d are forced to co-rotate with the vortex. For an external observer, these waves acquire negative energy \cite{patrick2024primer}, and their absorption by the vortex leads to superradiant amplification, fuelling the black-hole bomb. We illustrate this principle using the mode highlighted in Fig.~\ref{fig:experiment}d by a sinusoidal line. If the negative-energy wave (red) is absorbed inside the vortex, the wave outside gets amplified due to energy conservation. Formally, this wave undergoes repeated reflection at point B with reflectivity $|\mathcal{R}| > 1$, and its amplitude is expected to grow exponentially at a rate $\Gamma = \log(|\mathcal{R}|)/\tau$, where $\tau$ is the round-trip time between points B and C.

The parameter $\tau$ is determined by the shape of the potential barrier and the behaviour at the boundary (employing Neumann boundary conditions is experimentally justified in superfluid helium \cite{barroso2025holography}). Conversely, determining $\mathcal{R}$ requires a detailed understanding of wave behaviour at point A, which is typically located near the vortex core and outside the accessible field of view. In our superfluid system, quantum mechanics constrains circulation $C$ to multiples of the circulation quantum, $\kappa = 9.98 \times 10^{-8}~\mathrm{m^2\,s^{-1}}$ \cite{barenghi2023}. These quanta are carried by quantum vortices, line-like objects that naturally occur in superfluid helium. For our flow field, $N_C = 2\pi C/\kappa \approx 121{,}000$ vortices contribute to the central vortex, whose hollow funnel acts as a multiply quantised, or giant, vortex \cite{alperin2021}. Additional vortices distribute around it, forming a vortex core that may contain finite vorticity at scales larger than the mean separation between the vortex lines~\cite{inui2020,ruffenach2023}. Oscillations of this vorticity field may also lead to an instability~\cite{jansson2006polygons} and hence contribute to $|\mathcal{R}| > 1$. Modelling these effects in a dispersive, three-dimensional system lies beyond the scope of this work. However, in the simpler case of shallow-water waves around a Rankine vortex, we found that contribution of vorticity to amplification is subdominant compared to that of the ergoregion (and hence the black-hole bomb mechanism) when the vortex has a hollow core \cite{patrick2025sloshing}, as in our experiment.

The WKB model suggests that the most efficient amplification occurs for bound states, i.e. modes resonantly trapped between the potential barrier and the boundary. In our experiment, the sloshing wave oscillates with frequency $\omega_0/(2\pi) = 3.47$~Hz, and appears as a single horizontal feature in the spectrum shown in Fig.~\ref{fig:experiment}d (Measured). Although the WKB method is expected to break down at low frequencies, we predict that the lowest-frequency bound-state oscillates at $3.97$~Hz, already in reasonable agreement with the data in view of the approximations made. More importantly, the method reveals that the sloshing-mode frequency is determined by the properties of the vortex near the periphery of the flow. In this region, a more accurate model accounting for wave nonlinearities is required, as we discuss below.

\section*{Nonlinear one-mode-dominated dynamics}

The space- and time-resolved detection of interface fluctuations permits us to isolate individual modes by selecting a single $m$-band and applying a bandpass frequency filter within (see Methods). This approach yields time- and radially-resolved amplitudes $A$ of these modes, labelled $(m,\omega)$. In Fig.~\ref{fig:experiment}e, we present their radial averages $\left\langle A \right\rangle$ as a function of time. The amplitude of the sloshing, or primary, mode $(1,\omega_0)$ (red line) dominates the system, and exhibits exponential growth---in line with the black-hole bomb mechanism---at a rate $\Gamma = (0.081\pm 0.001)~\mathrm{s^{-1}}$.

The growth phase lasts only until $t \approx 80$~s, at which point the amplitude saturates at $(3.06 \pm 0.05)~\mathrm{mm}$ and the system enters a non-equilibrium steady state. Similar growth and saturation are observed for multiples of the primary mode (see the remaining curves in Fig.~\ref{fig:experiment}e), with the growth rate approximately doubling for $(2,2\omega_0)$ and further increasing for higher harmonics. This process is strikingly similar to the observations in \cite{gregory2024tracking}, where a one-mode reduced model with quartic self-coupling (neglecting feedback from other modes) quantitatively accounted for the amplitude saturation of a dominant mode excited by the Faraday instability~\cite{kumar1994parametric}. The recurrence of amplitude saturation in the present system suggests that nonlinear amplitude limitation may persist across distinct amplification mechanisms when the dynamics are organised around a dominant coherent mode. However, the wave decomposition in~\cite{gregory2024tracking} employs eigenfunctions of the homogeneous system, which cannot be transferred directly to the present vortex flow configuration.

\begin{figure}
    \centering
    \includegraphics[width=\columnwidth]{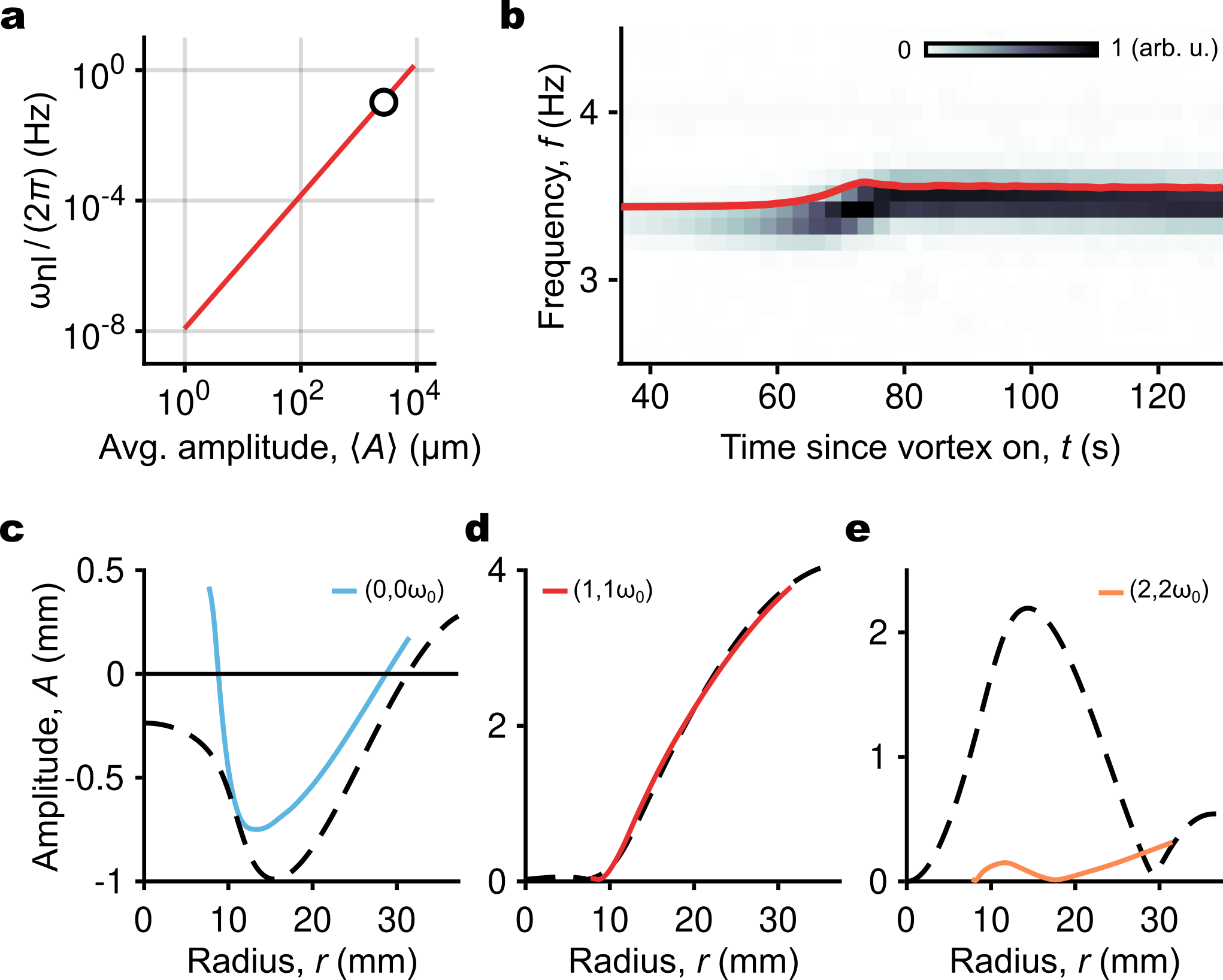}
    \caption{\textbf{Nonlinear wave evolution.} %
     \panel{a}~Red line denotes the calculated nonlinear frequency correction to $\omega_0$ as a function of the radially averaged amplitude $\left\langle A \right\rangle$ of the sloshing mode. The black point indicates the experimentally observed frequency drift of $0.098~\mathrm{Hz}$. %
    \panel{b} The expected drift of the sloshing frequency as a function of time (red line) matches the the spectrogram of the sloshing mode (colour map). %
    \panel{c-e}~Experimentally extracted mode profiles at $t = 65$~s (coloured lines) are compared to the predicted radial functions $A_0$, $A_1$ and $A_2$ (black dashed lines).%
    }
    \label{fig:model}
\end{figure}

We therefore adapt our modelling approach as follows. Given that the dominant contributions to the wave dynamics are the primary mode and its second harmonic, we expect the free surface deformation $\delta h$ to take the form
\begin{equation}
    \delta h(r,\theta,t) = A_1 e^{i(\theta - \omega_0 t)} + A_2 e^{i(2\theta - 2\omega_0 t)} + A_0 + \mathrm{c.c.},
     \label{eq:model}
\end{equation}
where c.c. denotes the complex conjugate. We factor out fast oscillations and leave the relatively slow exponential growth inside the amplitudes: $A_1(r,t)$ represents the primary mode $(1,\omega_0)$, $A_2(r,t)$ the second harmonic (which grows twice as fast as $A_1$, as confirmed by our experiment), and $A_0(r,t)$ quantifies the static deformation of the interface ($m=0,\omega=0$), a phenomenon known as backreaction in gravity simulators~\cite{patrick2021}. The primary mode drives the backreaction and second harmonic generation through nonlinearities in the fluid equations, which enter at higher order in perturbation theory when the wave amplitude is small relative to $h_0$.
Furthermore, if we restrict our attention to the periphery of the flow where the primary mode is localised, $v_\theta$ itself can be treated as a small correction to a fluid at rest, suggesting both the vortex and nonlinear wave mixing can be treated within a unified perturbative framework (see the Supplementary document for details).

Solving the equation of motion for each perturbation order, we first obtain the oscillation frequency of the primary mode, $\omega_0/(2\pi) = 3.44~\mathrm{Hz}$, now in agreement with the experimentally determined value of $3.47~\mathrm{Hz}$. This frequency is nevertheless expected to evolve as the amplitude of the primary mode increases. The nonlinear correction $\omega_\mathrm{nl}$ scales with $\left\langle A \right\rangle^2$~\cite{miles1984} (as shown in Fig.~\ref{fig:model}a) and, for the present experimental configuration, becomes experimentally detectable. This is indeed confirmed in Fig.~\ref{fig:model}b, where we compare the expected time evolution, $[\omega_0 + \omega_\mathrm{nl}(t)]/(2\pi)$, derived from the measured amplitude of the primary mode (red line), with that mode's spectrogram (see Methods). Despite a small offset between the predicted and observed frequencies, the cumulative frequency drift of $0.098~\mathrm{Hz}$ is fully compatible with the model (see the black point in Fig.~\ref{fig:model}a).

To further validate our model, we present in Fig.~\ref{fig:model}c-e the radial profiles of the backreaction, primary, and secondary waves for $t=65$~s, which lies within the exponential growth phase. We find satisfactory agreement between experimentally extracted mode shapes (coloured lines) and the theoretical radial profiles of $A_0$ and $A_1$ (black dashed lines). The agreement is less good for the secondary mode and $A_2$ (panel c), which can be attributed to (i) the fact that a significant portion of the wave is located in the vortex core region, where the perturbative treatment of the vortex is expected to fail, (ii) the fluctuations of the radial shape of this mode in time (unlike the backreaction and primary modes, whose shape remains approximately constant for the duration of the exponential growth phase; see Fig.~\ref{supp:mode-evo}); and (iii) a fundamental limitation of our model that allows only the excitation of a single higher-frequency mode. In fact, Fig.~\ref{fig:experiment}e shows multiple secondary excitations that undergo similar growth and saturation. This behaviour reflects the well-documented phenomenon of nonlinear wave mixing, previously observed in superfluid systems like trapped ultracold gases \cite{ott2003} or driven Faraday waves on the surface of superfluid helium \cite{levchenko2017}. We analyse these effects in the following section.

\section*{Wave mixing}

The excitation of higher-order modes can be understood as the result of wave mixing of order $N = n+1$, where $n$ waves of type $(1, \omega_0)$ interact and produce a single wave of type $(n, n \times \omega_0)$. Such processes are, in general, not limited to the primary mode, and one might anticipate the emergence of additional interacting modes contributing to the rich fluctuation spectrum we observe. To evidence the presence of nonlinear interactions between waves, we quantify correlations between the measured excitations~\cite{gregory2024tracking,schweigler2017experimental}.

We first consider the correlation between a selection of $N$ interacting modes. This time-resolved quantity establishes the presence of the wave mixing processes under investigation, and is calculated by applying the short-time Fourier transform over the time domain and spatially averaging the products of the chosen mode amplitudes. This measure is nevertheless sensitive to lower-order effects, where not all $N$ modes participate simultaneously. To exclude these lower-order contributions, we compute only the connected part of the correlation function $g_N$, also known as the Ursell function~\cite{ursell1927evaluation,tran2017lieb}, and normalise this quantity by the total amount of correlations at the same order. For details on the computation and normalisation, see Methods.

\begin{figure}[tb]
    \centering
    \includegraphics[width=83mm]{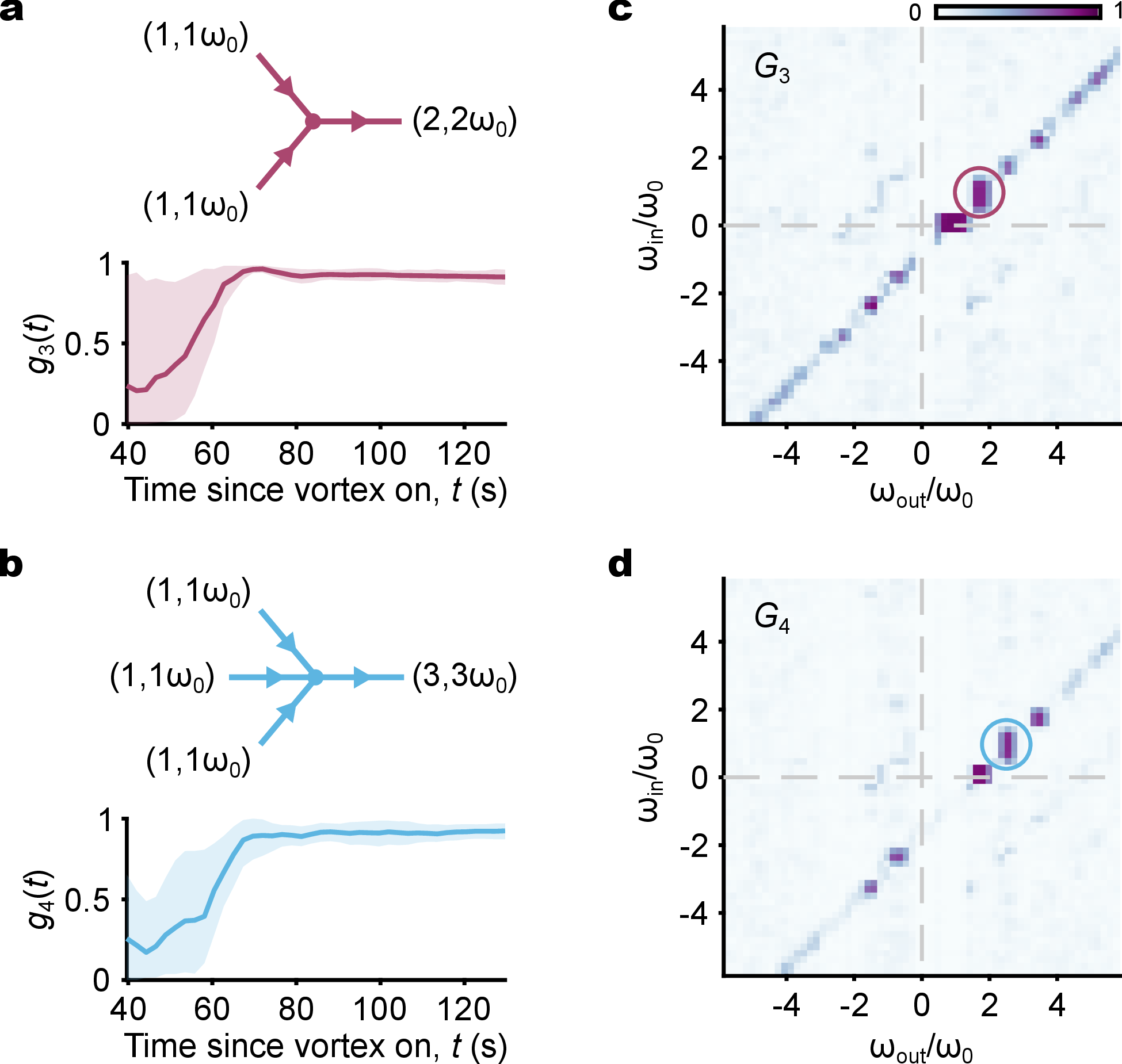}
    \caption{\textbf{Time- and frequency-resolved correlations.} %
    \panel{a,b} Time evolution of the connected correlation $g_N(t)$ for interaction orders $N = 3$ (a) and $4$ (b). Coloured lines show the mean, with shaded areas indicating $1\sigma$ confidence intervals over the real and imaginary components. The interaction diagrams above each plot illustrate the considered mixing processes: all incoming modes correspond to the primary instability $(1,\omega_0)$, while the outgoing modes represent the resulting excitations. %
    \panel{c,d} Frequency-resolved correlation $G_N$ for the interaction diagrams akin to those in panels a and b, with the varying frequency of the outgoing mode $\omega_\text{out}$ and of one incoming mode $\omega_\text{in}$. Each pixel aggregates contributions across all azimuthal numbers $m$. The colour scale represents the normalised interaction amplitude. Coloured circles mark the locations of processes to which panels a and b are contributing.}
    \label{fig:correlations}
    \vspace{-1em} 
\end{figure}

In Fig.~\ref{fig:correlations}a,b, we present the normalised, connected correlation functions $g_N(t)$ for three- and four-wave mixing of the primary mode, along with their corresponding interaction diagrams. Despite the relatively large uncertainty at early times (shaded area) due to a low signal-to-noise ratio at the beginning of the experimental run, both quantities clearly saturate near unity when the non-equilibrium steady state is established. Within our chosen normalisation, the saturation indicates that wave mixing at orders $N=3$ and $4$ is well resolved during the late stage of the black-hole bomb.

The presence of these specific processes in the data prompted us to explore a broader landscape of wave interactions by calculating the spectral connected correlation $G_N$ of the same orders (see Methods). Unlike $g_N$, this quantity incorporates correlations from all $m$-channels. The outgoing frequency $\omega_\text{out}$, as well as one of the incoming frequencies $\omega_\text{in}$ are treated as free parameters, while the remaining components are fixed to $\omega_0$, as these interactions are expected to dominate over those involving higher frequency components. The resulting correlation maps are shown in Fig.~\ref{fig:correlations}c,d. The dark features, which indicate more strongly correlated modes, follow lines where frequency, and thus energy, is conserved: $\omega_\text{out} = \omega_\text{in} + (N-2)\omega_0$. This suggests that frequency-preserving mode mixing of orders $N=3$ and $4$ is the dominant nonlinear effect, with higher-order processes presenting weaker correlation structures (see Fig.~\ref{supp:corr56}).

The self-interaction diagrams from panels a and b contribute to the single dark spots marked by violet and blue circles in panels c and d. However, the spectral correlations reveal the presence of nonlinear interactions extending across a broader range of frequencies, beyond simple multiples of $\omega_0$. These interactions may facilitate the redistribution of energy from the driven mode to a spectrum of secondary excitations, spanning multiple azimuthal numbers and frequencies. Consequently, the end state of the black-hole bomb is shaped by both the self-interaction of the unstable mode, initially subject to unbounded amplification, and a multitude of energy-conserving wave mixing processes.

We hypothesise that nonlinear interactions, which drive the excitation of secondary modes, also provide the primary mode with a critical energy-drain mechanism, thereby contributing to the stabilisation of the observed steady state. While a detailed examination of the physical processes underlying this mechanism lies beyond the scope of this work, our findings already underscore the pivotal role of nonlinearities in systems undergoing repeated superradiant amplification.

\section*{Conclusion}

The nonlinear effects investigated in this work are particularly timely and relevant to ongoing research in theoretical astrophysics. For example, a similar instability is expected for ultra-light axion fields surrounding a rotating black hole, where a low-frequency cloud forms and extracts energy and angular momentum through superradiance. This process generates distinctive gravitational-wave signals, making such systems targets for current and future detectors, and motivating the use of rotating black holes as powerful probes of physics beyond the Standard Model~\cite{Brito:2015oca,brito2017gravitational}. However, as in our experiment, the instability's end state is highly sensitive to the field's self-coupling. Particle production can saturate the instability before significant energy extraction occurs~\cite{fukuda2020aspects}, while attractive self-couplings may trigger a dramatic collapse of the axion cloud back into the black hole, reabsorbing its energy and angular momentum~\cite{yoshino2012bosenova}. While the saturation regime aligns with our results, the collapse scenario is echoed in cold atomic gas analogues of the black-hole bomb, where the instability instead leads to angular momentum accumulation at the drain~\cite{patrick2024mqv}. Nonlinear gravity simulators based on quantum fluids hence appear to be uniquely positioned to offer future insights and experimental feedback into how nonlinear phenomena reshape the observational signatures of real black holes~\cite{lehner2026}.

\vspace{2.5ex plus 1ex minus .2ex}

\paragraph*{\bf Acknowledgements}
P\v{S}, PS, VSB, SP, CFB, RG and SW extend their appreciation to the Science and Technology Facilities Council for their generous support within Quantum Simulators for Fundamental Physics (ST/T006900/1, ST/T005858/1, and ST/T00584X/1), as part of the UKRI Quantum Technologies for Fundamental Physics programme. LS, MTJ and SW gratefully acknowledge the support of the Leverhulme Research Leadership Award (RL-2019-020). MR acknowledges partial support from the Conselho Nacional de Desenvolvimento Científico e Tecnológico (CNPq, Brazil, grant 315991/2023-2), and from the São Paulo Research Foundation (FAPESP, Brazil, grants \mbox{2022/08335-0}, \mbox{2024/00923-6} and \mbox{2025/02701-3}). S.P. acknowledges support from the Engineering and Physical Sciences Research Council through the Stephen Hawking Postdoctoral Fellowship (EP/Z536660/1). SW also acknowledges the Royal Society University Research Fellowship (UF120112). RG and SW acknowledge support from the Perimeter Institute Research at Perimeter Institute is supported by the Government of Canada through the Department of Innovation, Science and Economic Development Canada and by the Province of Ontario through the Ministry of Research, Innovation and Science.

\paragraph*{\bf Author contributions} Experimental design and data acquisition by P\v{S}, LS, and PS. Data analysis by P\v{S}, PS, SS, MTJ, and VSB. Theoretical models developed by SP, LS, SG, and MR. Manuscript written by P\v{S}, with contributions from LS, SP, SS, MTJ, MR, and SW. Funding acquisition and manuscript revision by SW, RG, and CFB. Conceptualisation and project supervision by AA and SW.

\paragraph*{\bf Data \& Code availability} The datasets generated and analysed during this study are available upon reasonable request. This study does not rely on custom code or algorithms beyond standard numerical evaluations.

\paragraph*{\bf Materials \& Correspondence} Any correspondence or requests for data should be addressed to the corresponding author, P\v{S}.

\interlinepenalty=10000
\bibliography{biblio}

\clearpage\raggedbottom\newpage

\interlinepenalty=0

\section*{\large{Methods}}

\subsection{Experimental parameters}

The draining vortex is established above a circular drain with a diameter of 10 mm, housed within a cylindrical container of diameter $2r_B = 74.6$ mm. This container acts as a reflective boundary for surface waves. The vortex flow is sustained by continuously draining the superfluid through the central opening while refilling the experimental zone near the boundaries. This recirculation loop is maintained by a spinning 8-blade propeller positioned beneath the drain, which functions as a centrifugal pump \cite{obara2021}. The propeller is driven by a magnetically coupled DC motor operating at 240 rotations per minute (4~Hz).

The experimental cell is partially submerged in a superfluid bath, maintained at a constant temperature of $(1.70 \pm 0.01)$~K within a transparent glass cryostat. Thermal insulation, essential for preserving helium in its superfluid phase, is achieved through vacuum insulation and a liquid nitrogen jacket. We actively regulate the temperature by pumping helium vapours through a computer-controlled butterfly valve. The temperature stability results from balancing the parasitic heat flux into the experiment, the pumping rate, and the heat dissipated in the experimental zone. In the current setup, we typically observe an evaporation rate of $15~\mathrm{\mu m/s}$, equivalent to a 1.6-mm level drop over the duration of an experimental run, and corresponds to a cooling power of approximately 500~mW.

We image the free surface of superfluid helium with a high-speed digital camera (Phantom VEO-640L), capturing 21,400 frames at a rate of 200 frames per second and resolution of 1,536$\,\times\,$1,536 pixels ($52.6~\mathrm{\mu m\,pix^{-1}}$). The surface is illuminated against a periodic sinusoidal pattern (a high-resolution digital print on a transparent foil). The interface height fluctuation field, $\delta h$, is extracted from the recordings of the pattern’s deformations using synthetic Schlieren imaging \cite{moisy2009,wildeman2018}. The set-up is optimised for this purpose, and achieves sensitivity, defined as the smallest resolvable height difference between adjacent pixels, of approximately $0.5~\mathrm{\mu m}$. The captured pattern spans approximately 1.1 million pixels, but we downsample and polar-transform the height fluctuation field into 256 radial ($r$) and 128 azimuthal ($\theta$) samples for the analysis in polar coordinates.

The best fit of Eq.~\eqref{eq:velocity} to the azimuthal velocity shown in Fig.~\ref{fig:experiment}c yields circulation $C = (19.2 \pm 0.1)\times 10^{-4}~\mathrm{m^2\,s^{-1}}$ and solid-body rotation frequency $\Omega = (0.10 \pm 0.02)~\mathrm{rad\,s^{-1}}$. The number of quantum vortices equivalent to circulation $C$ is then $N_C = 2\pi C/\kappa = (121.1 \pm 0.5)\times 10^3$.

\subsection{WKB approximation}

The Wentzel–Kramers–Brillouin (WKB) approximation \cite{buhler2014} treats the wave dynamics in terms of the perturbed velocity potential $\delta\phi$, which is related to the fluctuations of the interface height through the kinematic relation $\partial_z\delta\phi = \partial_t\delta h$. Our modelling is based on \cite{patrick2024primer}. As stated in the main text, the fluctuations are considered to be plane waves with a spatially varying wave vector $\vect{k} = p \,\hat{\vect{r}} + (m/r) \,\hat{\vect{\theta}}$ and fixed angular frequency $\omega$. These waves follow the dispersion relation \cite{patrick2020},
\begin{align}
    \label{eq:disp}
    \left(\omega-\vect{v}\cdot\vect{k}\right)^2 = F(\vect{k})\,,
\end{align}
where the term $\vect{v}\cdot\vect{k}$ can be interpreted as a flow-induced blueshift. The function $F(\vect{k})$ describes dispersive properties of gravity-capillary waves, and reads
\begin{equation}
    \label{eq:dispfun}
    F(\vect{k}) = (gk + \gamma k^3)\tanh(h_0k)\,,
\end{equation}
where
$k \equiv ||\vect{k}||$, $g$ is the gravitational acceleration, and $\gamma = 2.22\times 10^{-6}$~m$^3$\,s$^{-2}$ is the ratio of surface tension and density \cite{donnelly1998}, evaluated for temperature at which our experiments took place.

For $m$ and $r$ fixed, we obtain a propagating wave solution for each real-valued radial wave vector $p$ that satisfies Eq.~\eqref{eq:disp}. We find these solutions as intersections of a line of constant $\omega$ and curves $\omega_D^\pm$, where
\begin{align}
    \omega_D^\pm(p,m,r) = \underbrace{\frac{mC}{r^2} + m\Omega}_{\vect{v}\cdot\vect{k}}\, \pm \, \sqrt{F(p,m,r)}\,.
    \label{eq:branch}
\end{align}
When a line of constant $\omega$ (see the solid blue line in Fig.~\ref{supp:dispsol}) crosses the upper branch $\omega_D^+$, we obtain two propagating waves: a radially ingoing one with $p<0$ and a radially outgoing with $p>0$. Similarly, we obtain a pair of negative-frequency (i.e. counter-rotating) waves as intersections of the $-\omega$ line with the lower branch $\omega_D^-$.

Note that the two branches display a gap at $p=0$. The frequency range inside the gap only contains evanescent solutions (dot-dashed yellow lines in Fig.~\ref{supp:dispsol}, which have no intersections with $\omega_D^\pm$). The size of this gap varies with $m$ and $r$, such that the curves $\omega_D^\pm(p=0,m,r)$ delimit the region of evanescent solutions in the frequency-radius diagram. This region is highlighted in Fig.~\ref{fig:experiment}d by violet shading.

As discussed in the main text, the region delimited by $\omega_D^+(p=0,m,r)$ and the reflective wall at the system's boundary forms a cavity that supports resonant bound states. The resonance condition reads
\begin{align}
    \label{eq:rescond}
    \int_{r_1}^{r_B} p(r)dr = \pi\left(n + \frac{1}{4}\right)\,,
\end{align}
where $r_1$ is the turning point (e.g. point B in Fig.~\ref{fig:experiment}d), and $n$ is a non-negative integer. The factor $1/4$ on the right-hand side is due to the waves' behaviour at $r_1$. We refer the readers to \cite{smaniotto2026} for a detailed description of scattering processes within the WKB approximation.

\subsection{Nonlinear perturbative model}

The perturbative model leading to mode profiles and the nonlinear frequency correction shown in Fig.~\ref{fig:model} is derived in the Supplementary document.

To compare experimental and theoretical mode profiles in Fig.~\ref{fig:model}c-e, we first match the amplitude of the primary mode ($1,1\omega_0$) by a least squares regression, then rescale the backreaction and secondary modes to the primary amplitude squared, as required by the model.

\subsection{Height fluctuation field}
\label{meth:c}

The height fluctuation field $\delta h(r,\theta,t)$ is analysed in Fourier space by applying the Fourier transform $\mathcal{F}$ to the azimuthal coordinate $\theta$ and time $t$. This yields the corresponding transformed coordinates $m$ and $f$, together with a complex-valued field $\xi(r,m,f)$ describing mode amplitudes. The colour map in Fig.~\ref{fig:experiment}d displays $|\xi(r,m=1,f)|$, rescaled to a unit interval for visualisation.

The amplitudes of specific modes are extracted by inverse-transforming the field $\xi$:
\begin{equation} \label{eqn:A_definition}
    A(r,t) = \left\vert \mathcal{F}^{-1} \left[ \xi(r,m=1,f) H(f) \right]\right\vert\,,
\end{equation}
where $H(f)$ is a raised-cosine bandpass filter, defined as
\begin{equation}
H(f) =
\begin{cases}
1 + \cos\left[\frac{\pi\left(f-f_0\right)}{w}\right] & \text{for } f \in \left(f_0-w, f_0+w\right), \\
0 & \text{elsewhere.}
\end{cases}
\label{eq:cos}
\end{equation}
We set $w = 0.5$~Hz. In Fig.~\ref{fig:experiment}e, we plot $\left\langle A \right\rangle$, the amplitude that is radially averaged within the available field of view, $r\in(7.0, 31.5)~\mathrm{mm}$. In Fig.~\ref{fig:model}c-e, we plot $A(r)$ for the given time instant and in Fig.~\ref{supp:mode-evo}, we show how these amplitudes vary with time.

The spectrogram shown in Fig.~\ref{fig:model}b is calculated as $\left\langle\left\vert \chi(r,m=1,f,t) \right\vert\right\rangle_r$, where $\chi$ is the short-time Fourier transform of $\delta h$. We use a sliding time window of 2,048 frames (10.24~s) with 50\% overlap and a symmetric Hann window function.

\subsection{Connected correlation functions}

The correlation between specific modes is derived from the short-time Fourier-transformed field $\chi(r, m, f, t)$, following the parametrisation outlined in the previous section. To quantify the accumulation of wave mixing, we adopt a field-theory approach \cite{gregory2024tracking} and define
\begin{equation}
    g^\text{full}_N(\chi_{1},\dots, \chi_{N}) = \langle \chi_{1}\dots \chi_{N}\rangle_{r,\theta},
    \label{eq:gnfull}
\end{equation}
where the subscript $r,\theta$ denotes spatial averaging over these coordinates. Term $\chi_i$ represents the mode amplitude $\chi$ after selecting the azimuthal number $m_i$ and frequency $f_i$, then transforming back to direct space. This leaves $\chi_i$ a function of $r$, $\theta$, and $t$. The correlation function $g^\text{full}_N$ encompasses information on all coherent wave mixing processes involving up to $N$ participating modes. To isolate contributions where all $N$ modes interact simultaneously, we focus solely on the connected component of the correlation function, $g^\text{con}_N$. This is achieved by subtracting lower-order contributions \cite{schweigler2017experimental} through the iterative procedure described in \cite{gardiner2004handbook}.

The correlation functions presented in Fig.~\ref{fig:correlations}a,b are calculated as
\begin{equation}\label{eqn:g_N}
    g_N(t) = \frac{|g^\text{con}_N (\chi_1 \dots \chi_N)|}{\left\langle | \chi_1 \dots \chi_N | \right\rangle_{r,\theta}}\,.
\end{equation}
This normalisation allows us to assess the relative importance of fully connected contributions in comparison to all correlations of order $N$. As detailed in the main text, our analysis demonstrates that the observed wave mixing of orders 3 and 4 primarily accumulate correlations through fully connected processes. We have additionally evaluated other widely used normalisations, including standardised cumulants \cite{gardiner2004handbook}, and found no evidence of anomalous behaviour.

Frequency-resolved correlation functions are defined similarly to Eq.~\eqref{eq:gnfull}, as
\begin{equation}
\label{eqn:G_N}
    G_N^\text{full} (\omega_1,\dots,\omega_N) = \langle \chi_{\omega_1}\dots \chi_{\omega_N}\rangle_{r,\theta, t}\,,
\end{equation}
where the averaging is performed over all spatial dimensions, $r$, $\theta$, and $t$. The modes $\chi_{\omega_i}$ are obtained by selecting only the frequency $f_i = \omega_i/(2\pi)$ from the Fourier-transformed field $\chi$, leaving $\chi_{\omega_i}$ a function of $r$, $\theta$, and $t$. Following the same procedure used to calculate $g_N$, we subtract lower-order contributions to isolate the connected component, $G_N^\text{con}$. We then normalise this quantity to obtain the frequency-resolved correlation, $G_N$, given by
\begin{equation}
    G_N = \frac{|G_N^{\text{con}}(\omega_1,\dots,\omega_N)|}{\langle|\chi_{\omega_1}\dots \chi_{\omega_N}|\rangle_{r,\theta, t}}\,,
\end{equation}
The correlation maps in Fig.~\ref{fig:correlations}c,d are generated using two free parameters, $\omega_1 = \omega_\text{in}$ and $\omega_N = \omega_\text{out}$, while the remaining frequencies $\omega_i$ are fixed to $\omega_0$, the frequency of the primary instability.

Higher-order correlators can be computed at the expense of increased noise. As shown in Fig.~\ref{supp:corr56}, the correlation maps for $N=5$ and $6$ still exhibit signatures of frequency-conserving processes. However, these signals are embedded within a relatively noisy background.

Finally, we confirmed that all correlation functions presented in Fig.~\ref{fig:correlations} converge to within 5\% across all dimensions used for averaging. Additionally, we conducted a bootstrapping error analysis on the dataset, which revealed strong consistency in all reported results, with errors at the level of the 5\% convergence threshold. However, due to the relatively limited size of the dataset, a detailed assessment of statistical significance could not be achieved.

\clearpage\raggedbottom\newpage

\onecolumngrid
\section*{\large{Supplementary data}}

\setcounter{figure}{0}
\renewcommand{\thefigure}{S\arabic{figure}}

\begin{figure*}[h!]
    \centering
    \includegraphics[width=180mm]{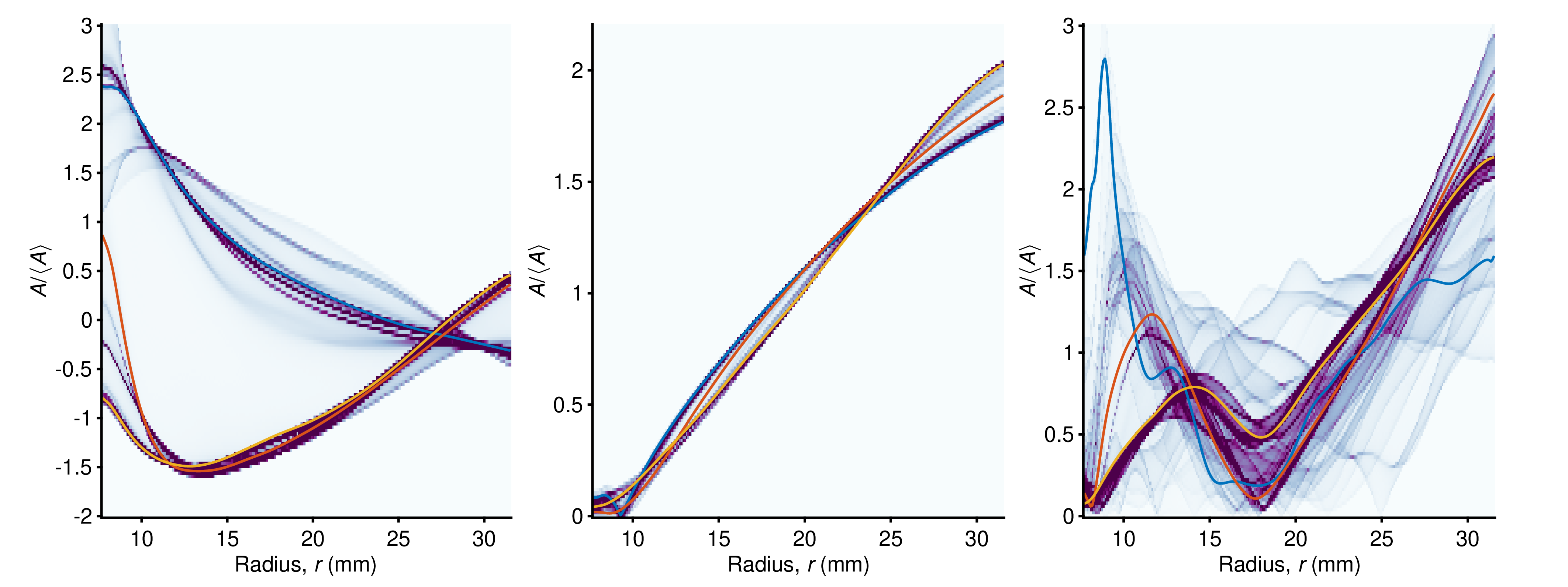}
    \caption{\textbf{Time evolution of dominant modes.} Radial shapes of the backreaction (left), primary (middle) and secondary modes (right) are shown for $t = 40$~s (blue), $65$~s (red) and $90$~s (yellow lines). The instantaneous amplitude $A$ is normalised by $\left\langle A \right\rangle$, which denotes the radial root mean square for the backreaction mode, and radial averages for the primary and secondary modes. To appreciate how the mode shapes vary in time, we plot by the blue-violet colour map the histograms of all mode profiles between $t = 30$ and $90$~s.}
    \label{supp:mode-evo}
    \vspace{-1.5em}
\end{figure*}

\begin{figure*}[h!]
    \centering
    \includegraphics[width=40mm]{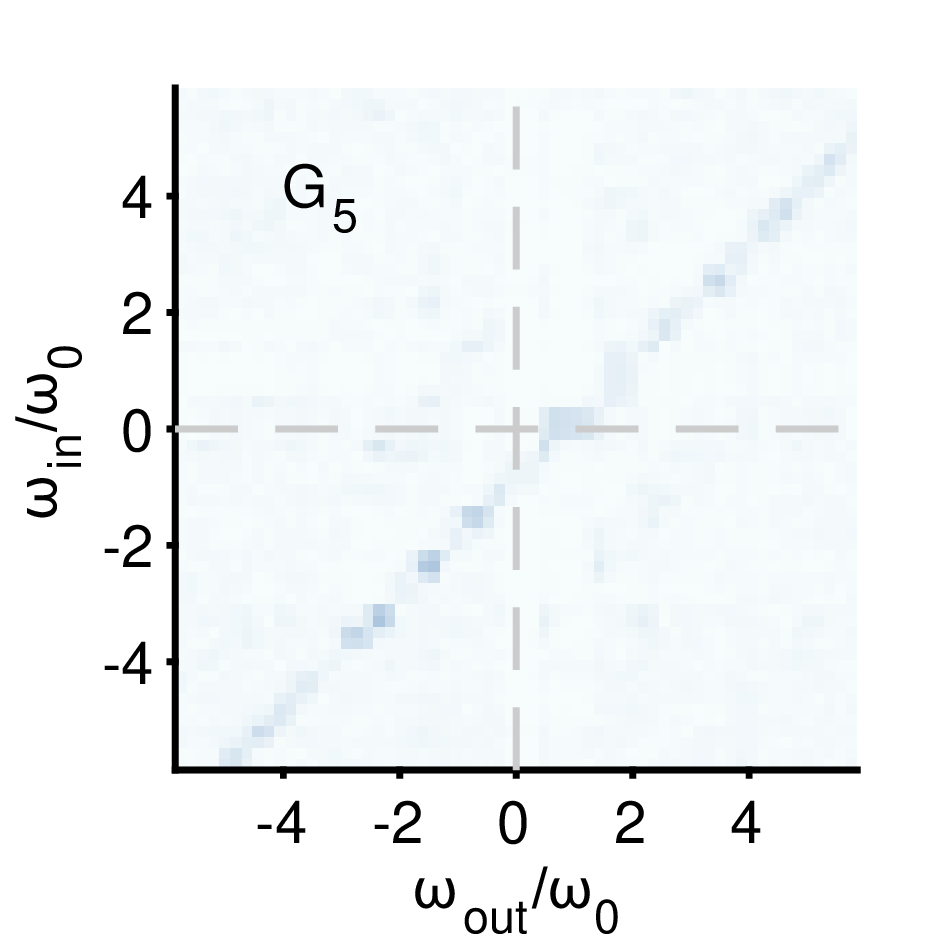} \includegraphics[width=40mm]{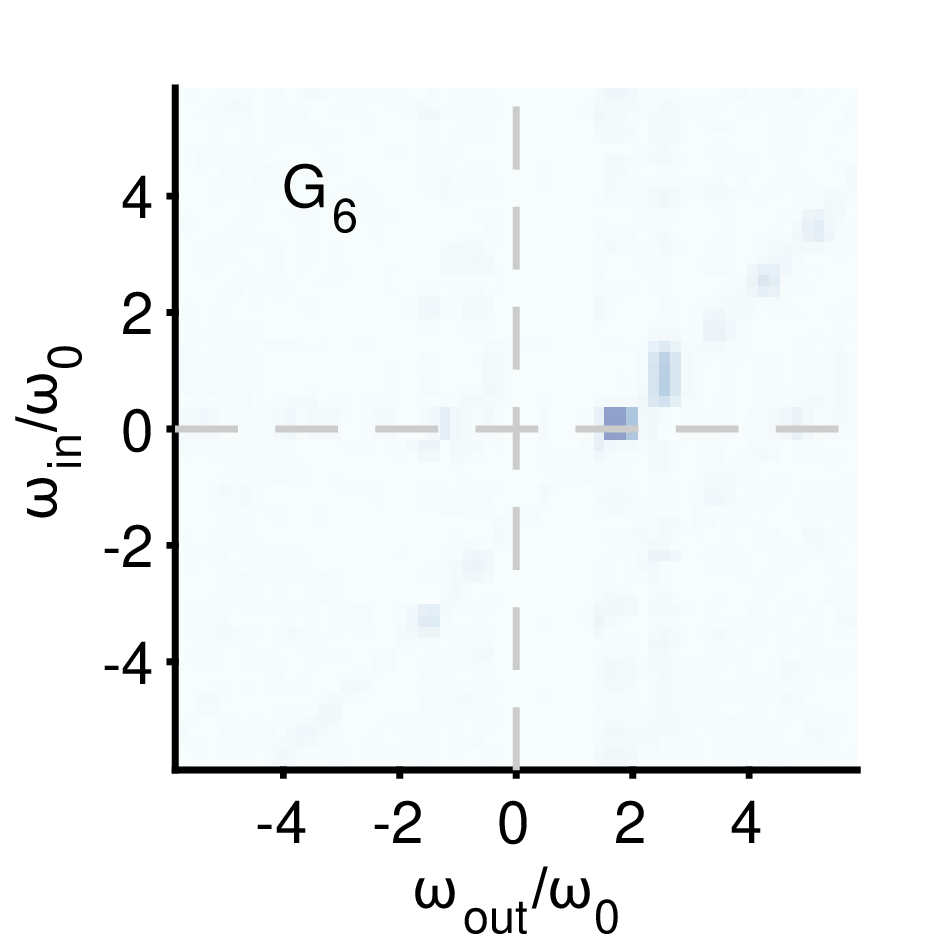}
    \caption{\textbf{Correlations of order $\bm{N=5,6}$.} The correlation functions are plotted with the same colour scale as $G_3$, $G_4$ in Fig.~\ref{fig:correlations} and indicate that correlations from 5- and 6-wave mixing processes are less pronounced compared to lower-order processes.}
    \label{supp:corr56}
    \vspace{-1.5em}
\end{figure*}

\begin{figure}[h!]
    \centering
    \includegraphics[width=80mm]{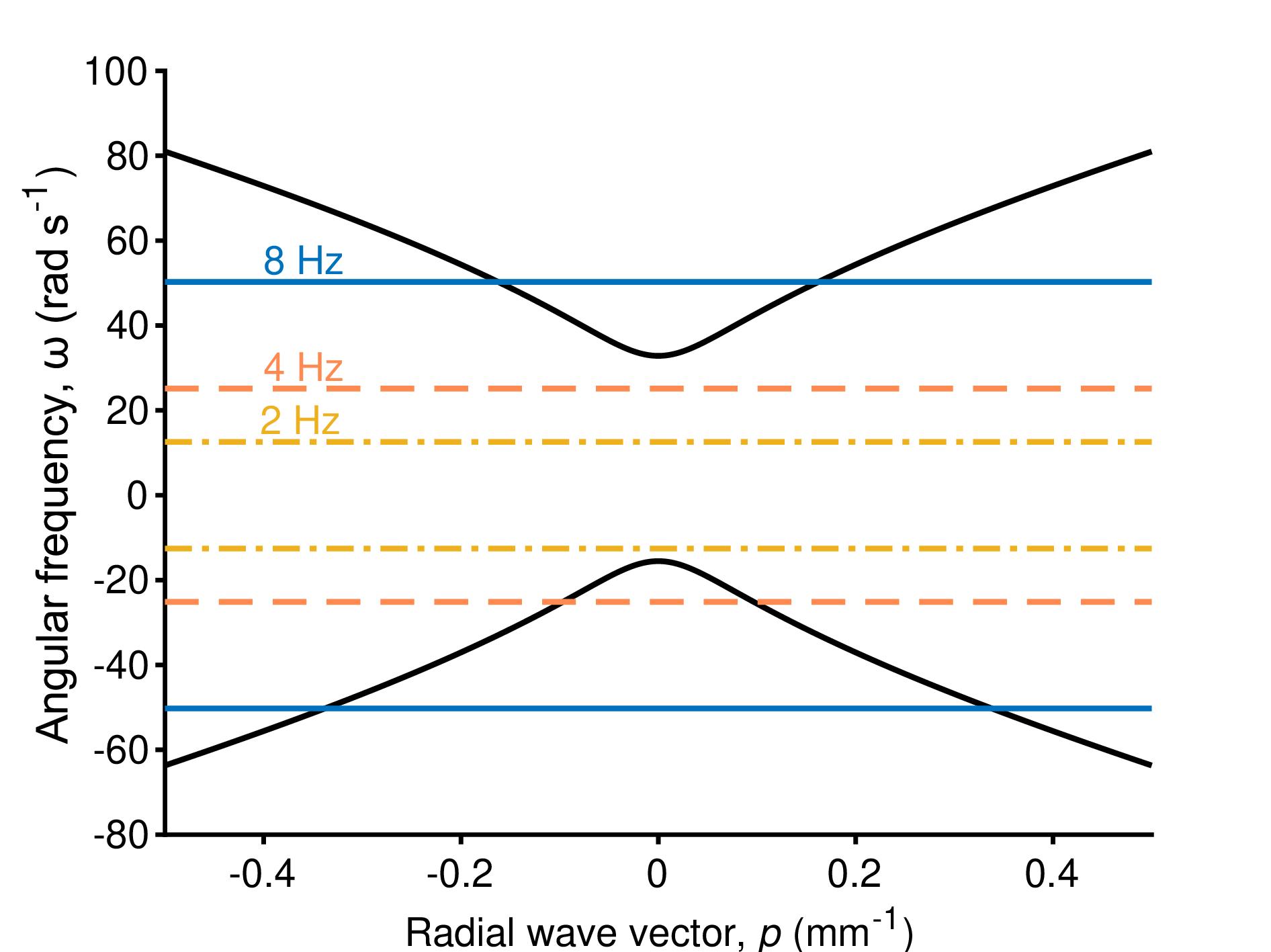}
    \caption{\textbf{Propagating and evanescent wave solutions.} Solid black curves indicate $\omega_D^+$ (in the $\omega > 0$ half-plane) and $\omega_D^-$ ($\omega < 0$ half-plane). Three pairs of lines with constant frequency represent 3 sets of wave solutions based on the number of intersections with $\omega_D^\pm$: evanescent waves only (dot-dashed yellow), propagating counter-rotating waves (dashed red), and propagating co- and counter-rotating waves (solid blue). When a constant line intersects one of the $\omega_D^\pm$ curves, two propagating solutions represent a radially outgoing ($p > 0$) and a radially ingoing ($p < 0$) wave. The flow parameters correspond to the experiment discussed in the main text, and we take $m=1$ and $r = 15$~mm.}
    \label{supp:dispsol}
\end{figure}

\clearpage\raggedbottom\newpage

\section*{\large{Supplementary document}}

\setcounter{equation}{0}
\renewcommand{\theequation}{S\arabic{equation}}

\section{Perturbative model of the vortex}

The WKB model presented in the main text provides an intuitive picture of wave propagation in our system, including the instability mechanism.
However, the approximation is best-suited to short wavelength excitations which oscillate over much shorter scales than variations in the background flow.
In contrast, the observed sloshing (primary) wave has a wavelength comparable to the system size, being the fundamental (lowest frequency) mode with $m=1$.
Hence, it is unsurprising that the discrepancy between the WKB prediction ($3.97$ Hz) and the measured frequency ($3.44$ Hz) is $17\%$.
Furthermore, the calculation shows that the growth rate depends sensitively on the details of the flow around the drain hole, which our detection method is unable to resolve, whereas the oscillation frequency is mainly determined by the flow near the boundary, where the sloshing wave is trapped.
In view of these considerations, we seek an alternative approximation scheme to model the oscillation frequency of the sloshing mode, restricting our attention to the periphery of the flow.
In this region, the velocity field and the free surface depression caused by the vortex can be treated perturbatively.
This framework allows naturally for the inclusion of nonlinear effects in the wave amplitude, which Fig. 2 shows are important at late times in our system.

The key step in the argument is the identification of a suitably small parameter $\epsilon$, which we use to expand the equations of motion perturbatively.
We denote the angular velocity field $v_\theta$, the asymptotic surface height $h_0$, surface excitations $\eta$, the free surface depression $\Delta h$, and $\omega_1$ and $\Gamma_1$ the primary mode oscillation frequency and growth rate.
Near the outer cylinder wall, where the velocity field is small, we have $v_\theta/c\sim 0.1$, where $c=\sqrt{gh_0}$ (i.e. the shallow water wave speed) is the relevant characteristic speed for our system.
Hence, if we define $\epsilon \simeq 0.1$, we have the following hierarchy,
\begin{equation} \label{hierarchy}
\frac{v_\theta}{c},\frac{\eta}{h_0} \sim \epsilon, \qquad \frac{\Delta h}{h_0} \sim \epsilon^2, \qquad \frac{\Gamma}{\omega}\sim \epsilon^3,
\end{equation}
where for $\eta$ we take a representative value when the primary mode has attained its maximum amplitude.
This hierarchy determines at which order in perturbation theory different effects will enter, once we expand the equations of motion.

\subsection{Surface wave Lagrangian}

We start by deriving a set of equations for the nonlinear wave motion.
Figs. 3 and S2 demonstrate that 3 and 4 wave mixing induce the majority of the observed correlation structures, hence, we restrict ourselves to a Lagrangian that goes to fourth order in the wave amplitudes.
We initially consider a flat fluid interface, although a free surface depression can be introduced formally within the perturbation terms, as we do at the end of this section.
 
An irrotational, incompressible fluid with a free surface located at $z = h=\mathrm{const.}$ is governed by the following system of equations,
\begin{equation}
\begin{split}
    & \left[\dot{\Phi} + \frac{1}{2}(\grad\Phi)^2 +\frac{1}{2}(\partial_z\Phi)^2 + gh - \gamma\grad\cdot\left(\frac{\grad h}{\sqrt{1+|\grad h|^2}}\right)\right]_{z=h} = 0, \\
    & \partial_z\Phi|_{z=h} = \partial_t h + \grad h \cdot\grad\Phi|_{z=h},
\end{split}
\end{equation}
where $\grad$ acts in the $(x,y)$-plane.
The strategy is to perturb $h\to h+\eta$ and $\Phi\to\Phi+\phi$ and retain up to and including cubic terms in the equations of motion.
An expression for velocity potential perturbations can be found by integrating Laplace's equation from the impermeable lower boundary up to a height $z$,
\begin{equation}
    \phi = \cosh(-iz\grad)\phi_0 = \sum_{n=0}^\infty \frac{(-iz\grad)^{2n}}{(2n)!}\phi_0.
\end{equation}
We expand the following quantities,
\begin{equation}
\begin{split}
    \dot{\phi}|_{z=h+\eta} = & \ \dot{\phi}_h + \eta F\dot{\phi}_h - \frac{\eta^2}{2}\nabla^2\dot{\phi}_h + \mathcal{O}(\eta^3\phi_0), \\
    \grad\phi|_{z=h+\eta} = & \ \grad\phi_h + \eta \grad F\phi_h - \frac{\eta^2}{2}\nabla^2\grad\phi_h + \mathcal{O}(\eta^3\phi_0), \\
    \partial_z\phi|_{z=h+\eta} = & \ F\phi_h -\eta\nabla^2\phi_h - \frac{\eta^2}{2}\nabla^2 F\phi_h + \mathcal{O}(\eta^3\phi_0),
\end{split}
\end{equation}
where $\phi_h = \phi(z=h)$ and $F(-i\grad) = -i\grad\tanh(-ih\grad)$ and we have used the constancy of $h$ to commute $F$ through $\grad$.
Inserting into the Bernoulli equation and assuming the background flow to be independent of $z$,
\begin{equation}
\begin{split}
    & D_t\phi + g\eta - \gamma\nabla^2\eta + \\
    + & \, \eta D_t F\phi + \frac{1}{2}|\grad \phi|^2 + \frac{1}{2}(F\phi)^2 + \\
    + & \, \eta\grad\phi\cdot\grad F\phi - \eta(F\phi)\nabla^2\phi + \frac{\gamma}{4}\grad\cdot\left(|\grad\eta|^2\grad\eta\right) + \\
    + & \, \text{quartic terms} = 0,
\end{split}
\end{equation}
where $D_t = \partial_t+\vect{v}\cdot\grad$ and we have grouped the different orders line by line. We also dropped the subscript on $\phi_h$ and assume all quantities are evaluated at $z=h$ (i.e. we have an effective theory at the free surface).
The height equation is,
\begin{equation}
    D_t\eta - F\phi + \grad\cdot\left(\eta\grad\phi + \frac{\eta^2}{2}\grad F\phi\right) +  \, \text{quartic terms} = 0.
\end{equation}
Working to the same order of accuracy, we can manipulate these equations by substituting for $F\phi$ and dropping extra quartic terms,
\begin{equation} \label{nonlinEOM}
\begin{split}
    & D_t\eta + F\phi + \grad\cdot\left(\eta\grad\phi + \frac{\eta^2}{2}\grad D_t\eta\right) = 0, \\
    & D_t\phi + g\eta - \gamma\nabla^2\eta + \frac{1}{2}|\grad\phi|^2 + \eta D_t^2\eta + \frac{1}{2}(D_t\eta)^2 + \\
    & + \frac{\eta^2}{2}D_t\nabla^2\phi + \grad\cdot\left(\eta D_t\eta\grad\phi\right) + \eta D_t(\grad\eta\cdot\grad\phi) + \frac{\gamma}{2}\grad\cdot\left(|\grad\eta|^2\grad\eta\right) = 0.
\end{split}
\end{equation}
The benefit of writing the equations this way is that they can be derived from the following action,
\begin{equation} \label{action}
\begin{split}
    S = & \, \rho\int d^2\vect{x} \bigg\{-\eta D_t\phi - \frac{1}{2}\phi F\phi - \frac{g}{2}\eta^2 - \frac{\gamma}{2}|\grad\eta|^2 -\frac{1}{2}\eta|\grad\phi|^2 + \frac{1}{2}\eta(D_t\eta)^2 \\
    & \, -\frac{1}{6}\eta^3 D_t\nabla^2\phi + \eta D_t\eta\grad \eta\cdot\grad\phi + \frac{\gamma}{8}|\grad\eta|^4\bigg\},
\end{split}
\end{equation}
which is valid up to and including quartic interactions.
If we want to include a small background surface deformation $\Delta h(\vect{x})$, we can make the replacement $\eta\to \eta + \Delta h(\vect{x})$ provided $\Delta h$ is at least as small as the waves $(\phi,\eta)$.
The action \eqref{action} forms the basis of the following analysis.

\subsection{Linear theory}

Before including nonlinear effects, we first apply a perturbative treatment of the vortex to the linear wave theory.
Starting from the surface wave action in Eq.~\eqref{action}, we obtain the linear equations,
\begin{equation} \label{linear}
\begin{split}
D_t\eta + 2\Gamma_1\eta - F\phi + \grad\cdot(\Delta h \grad \phi) = & \ 0, \\
D_t\phi + g\eta - \gamma\nabla^2\eta + \Delta h D_t^2\eta = & \ 0,
\end{split}
\end{equation}
where $\Gamma_1$ is a phenomenological parameter which needs to be added by hand to compensate the fact that our assumption of weak $v_\theta$ ultimately neglects dissipation inside the vortex core.

Next, we decompose the waves into the different ($m,\omega$) components which, at linear order, amounts to setting $\phi = \phi_{mn} e^{im\theta-i\omega_{mn} t} + \mathrm{c.c.}$ and similarly for $\eta$, where $n$ is used to index a particular mode of the field (e.g. $n=0$ is the lowest frequency, fundamental mode).
Defining the state vector $|\psi_j\rangle = (\phi_j, \eta_j)^\mathrm{T}$ where $j=\{m,n\}$, we rewrite Eqs.~\eqref{linear} as,
\begin{equation} \label{linear2}
\omega_j|\psi_j\rangle = (L+i \epsilon^3 D)|\psi_j\rangle, \qquad L = L_{0,m} + \epsilon L_{1,m} + \epsilon^2 L_{2,j},
\end{equation}
where $L$ is a linear operator whose components are given by,
\begin{equation}
\begin{split}
L_{0,m} = & \ -i\begin{pmatrix}
0 & T_m \\ -F_m & 0
\end{pmatrix}, \qquad \epsilon L_{1,m} = \frac{m v_\theta}{r}\begin{pmatrix}
1 & 0 \\ 0 & 1
\end{pmatrix}, \quad\text{and}\quad
\epsilon^2 L_{2,j} = \ -i\begin{pmatrix}
0 & -\Delta h \,\tilde{\omega}_j^2 \\ \grad_m\cdot\Delta h\grad_m & 0
\end{pmatrix},
\end{split}
\end{equation}
with $\grad_m = (\partial_r, im/r)$, $F_m = F(-i\grad_m)$, $T_m = g -\gamma\nabla^2_m$ and $\tilde{\omega}_j = \omega_j - mv_\theta/r$.
The dissipation function is,
\begin{equation}
\epsilon^3 D = 2\Gamma_1\begin{pmatrix}
0 & 0 \\ 1 & 0
\end{pmatrix}.
\end{equation}
We also define the inner product,
\begin{equation}
\langle \psi_j | O \psi_j \rangle_y = \rho \int d^2\vect{x} \, (\phi^*_j \ \, \eta^*_j) \sigma_y O\begin{pmatrix}
\phi_j \\ \eta_j
\end{pmatrix},
\end{equation}
where $O$ is a general operator, $\sigma_y$ is the Pauli-$y$ matrix and the integral spans the domain $r\in[0,r_B]$.
The constant prefactor $\rho$ is included in the definition so that $\omega_j\mathcal{N}_j$ corresponds to the physical wave energy, where $\mathcal{N}_j = \langle\psi_j|\psi_j\rangle_y$ is the normalisation.

The benefit of this splitting is that we can separate the contributions at various orders in $\epsilon$.
We expand,
\begin{equation} \label{linear_ptb}
\begin{split}
\omega_j = & \ \omega^{(0)}_j + \epsilon \omega^{(1)}_j + \epsilon^2 \omega^{(2)}_j + \cdots \\
|\psi_j\rangle = & \ |\psi^{(0)}_j\rangle + \epsilon |\psi^{(1)}_j\rangle + \cdots
\end{split}
\end{equation}
where the powers of $\epsilon$ in $|\psi_j\rangle$ imply the order relative to the leading $|\psi^{(0)}_j\rangle$ term, which itself is formally $\mathcal{O}(\epsilon)$.
Working to this order in perturbation theory, we can neglect the dissipation term which enters for the first time at $\mathcal{O}(\epsilon^4)$ in the equations of motion.
This amounts to defining a fast time $t\sim\omega^{-1}_j$ on which oscillations occur, and a slow time $t'\sim\Gamma_1^{-1}$ over which the wave grows exponentially.
We restrict our attention to the fast dynamics and assume that $\omega_j$ and $|\psi_j\rangle$ adjust adiabatically over slow time.

The eigenfunctions of the vortex-free state are Bessel functions,
\begin{equation} \label{bessel}
\begin{split}
|\psi^{(0)}_j\rangle = & \ \alpha_j J_m(k_j r) \begin{pmatrix}
1 \\ i\omega_j^{(0)}(g+\gamma k_j^2)^{-1}
\end{pmatrix}, \\
\omega_j^{(0)} = & \ \frac{\langle\psi^{(0)}_j|L_{0,1}|\psi^{(0)}_j\rangle_y}{\mathcal{N}^{(0)}_j} = \sqrt{(gk_j+\gamma k_j^3)\tanh(k_j h_0)}
\end{split}
\end{equation}
where $\alpha_j\sim\mathcal{O}(\epsilon)$ is an arbitrary amplitude and the allowed $k_j$ are determined by the no-penetration boundary condition $\partial_r J_m(k_j r_B) = 0$.
Focussing on the primary mode with $j=\{1,0\}$ (i.e. the $m=1$ fundamental $n=0$ mode), higher order frequency corrections are determined by,
\begin{equation} \label{omega_lin}
\begin{split}
\omega_j^{(1)} = \frac{\langle\psi^{(0)}_j|L_{1,1}|\psi^{(0)}_j\rangle_y}{\mathcal{N}^{(0)}_j}, \qquad \omega_j^{(2)} = \frac{\langle\psi^{(0)}_j|L^{(0)}_{2,j}|\psi^{(0)}_j\rangle_y}{\mathcal{N}^{(0)}_j} + \sum_{j'}\frac{|\langle\psi^{(0)}_{j'}|L_{1,1}|\psi^{(0)}_j\rangle_y|^2}{\mathcal{N}^{(0)}_j\mathcal{N}^{(0)}_{j'} (\omega_j^{(0)}-\omega_{j'}^{(0)}) },
\end{split}
\end{equation}
where $j'=\{1,n'\}$ with $n'\neq 0$, and we added a superscript $(0)$ to the $L_2$ operator to indicate that it involves only the leading order contribution to the frequency, i.e.
\begin{equation}
\langle\psi^{(0)}_j|L^{(0)}_{2,j}|\psi^{(0)}_j\rangle_y = \rho\int d^2\vect{x}\,\Delta h\left(|\grad_1\phi^{(0)}_j|^2 - |\omega^{(0)}_j\eta^{(0)}_j|^2 \right).
\end{equation}
Since the first term is larger than the second, a free surface depression ($\Delta h< 0$) reduces the frequency and therefore the mode energy.
The first order correction to the eigenfunction is,
\begin{equation}
|\psi^{(1)}_j\rangle = \sum_{j'}\frac{\langle\psi^{(0)}_{j'}|L_{1,1}|\psi^{(0)}_j\rangle_y}{\mathcal{N}^{(0)}_{j'} (\omega_j^{(0)}-\omega_{j'}^{(0)}) }|\psi^{(0)}_{j'}\rangle
\end{equation}
There is a subtle point that our approach assumes $v_\theta$ to be small, however, the velocity model in Eq. (1) of the main text becomes infinite on the vortex axis.
We regulate this divergence by introducing a physically motivated model \cite{vatistas1991simpler},
\begin{equation} \label{velocity}
v_\theta(r) = \frac{Cr}{(a^{2q} + r^{2q})^{1/q}}, \qquad h(r) = h_0\left(1 - \frac{r_a^2}{r^2}\right).
\end{equation}
This lowers the velocity field in the low $r$ region, matching the decrease observed in Fig. 1.
For $q=6$, we find the best fit parameters $C = 19.7\,\mathrm{cm^2 s^{-1}}$ and $a = 1.1\,\mathrm{cm}$.
For the free surface depression, we take the model,
\begin{equation}
\Delta h(r) = -\frac{C^2}{2gr^2}\Theta(r-r_a) - h_0 \Theta(r_a-r).
\end{equation}
which is the formal solution for an irrotational vortex with $v_\theta = C/r$,
where $r_a = C/\sqrt{2gh_0}$ is the radius of the hollow core of the vortex.
Although the assumption $|\Delta h|\ll h_0$ is clearly violated for low $r$, the mode functions for $m\neq 0$ go to zero as $r\to 0$, hence, the integrals above for the higher order frequency corrections do not depend on what happens in this region.
Finally, we use Eq.~\eqref{linear_ptb} to compute $\omega_{1,0} = 2\pi\times 3.44\,\mathrm{Hz}$ for the primary mode in the linear regime.
The eigenfunction $\eta^{(0)}_{1,0} + \epsilon \eta^{(1)}_{1,0}$ is shown on Fig. 1d of the main text, and agrees remarkably well with the data.

\subsection{Nonlinear corrections}

Nonlinear corrections to the frequency in the wave amplitude are also captured at $\mathcal{O}(\epsilon^2)$ in perturbation theory.
These corrections arise in the following way: firstly, the primary $m=1$ mode drives higher order excitations in the $m=0,2$ bands at $m\times$ the primary frequency; secondly, these driven modes feed back into the equation for $m=1$ as subdominant corrections, leading to a shift in the primary mode frequency.
Since we assume that $m=1$ contains only the primary mode, we can simplify the notation by replacing the subscript $j$ with a $1$ (denoting the azimuthal index),
\begin{equation}
\begin{split}
 \omega^{(0)}_j + \epsilon \omega^{(1)}_j + \epsilon^2 \omega^{(2)}_j \to & \ \omega_1, \\
|\psi^{(0)}_j\rangle + \epsilon |\psi^{(1)}_j\rangle \to & \ |\psi_1\rangle.
\end{split}
\end{equation}
Then the total wave field including the driven $m=0,2$ modes is,
\begin{equation}
(\phi, \ \eta)^\mathrm{T} = |\psi_1\rangle e^{i(\theta -\omega_1 t)} + |\psi_2\rangle e^{2i(\theta-\omega_1 t)} + |\psi_0\rangle + \mathrm{c.c.},
\end{equation}
where $|\psi_{0,2}\rangle$ are solutions of,
\begin{equation}
\begin{split}
L_{0,0}|\psi_0\rangle  = & \ i\begin{pmatrix}
\frac{1}{2}\grad_1\phi_1\cdot\grad^*_1\phi^*_1 - \frac{1}{2}\tilde{\omega}_1^2|\eta_1|^2 \\[1.5ex] \mathrm{Re}[\grad_1^*\eta^*_1\cdot\grad_1\phi_1 +\eta_1^*\nabla^2_1\phi_1]
\end{pmatrix}, \\[3ex]
(L_{0,2}-2\tilde{\omega}_1 )|\psi_2\rangle = & \ i\begin{pmatrix}
\frac{1}{2}\grad_1\phi_1\cdot\grad_1\phi_1 - \frac{3}{2}\tilde{\omega}_1^2\eta_1^2 \\[1.5ex] \grad_1\cdot(\eta_1\grad_1\phi_1)
\end{pmatrix},
\end{split}
\end{equation}
and we defined $\tilde{\omega}_1 = \omega_1 - v_\theta/r$.
The $m=0$ equation has a simple solution. Using Eq.~\eqref{bessel} to evaluate the right hand side, we find that second equation for $m=0$ (lower entry) vanishes, and therefore we can set $\phi_0 = 0$.
Neglecting surface tension, which is a reasonable assumption since the perturbations of interest have a wavelength which is much larger than the capillary length $\sqrt{\gamma/g}$, we find,
\begin{equation}
\eta_0 \simeq -\frac{1}{2g}\grad_1\phi_1\cdot\grad^*_1\phi^*_1 + \frac{\tilde{\omega}_1^2}{2g}|\eta_1|^2,
\end{equation}
which encodes the backreaction onto the height function.
The equation for the secondary wave $\eta_2$ needs to be solved numerically.
We compare measurements of $\eta_{0,2}$ with the measured profiles in Fig. 2(c,e) of the main text.
The $m=0$ profile is consistent with the data.
For $m=2$ the agreement is less good, although the wave has roughly the correct amplitude near the edge of the flow when the perturbative analysis is expected to work best.
A possible reason for the large discrepancy is that the perturbative method predicts significant concentration of wave energy in the centre of the vortex where the velocity field is non-perturbative. 
The backreaction correction is not sensitive to this effect since there is no $m v_\theta/r$ term in the $m=0$ equations, explaining why the agreement with the data is much better in that case.

Next, we can calculate corrections to the primary mode frequency that are quadratic in the primary mode amplitude.
These enter at third order in the $m=1$ equation and can be split into three contributions: cubic interaction (at the level of the action) between two primaries and a secondary; cubic interactions between two primaries and the backreaction field; and a quartic interaction between four primaries.
Calculating these terms is much more involved than the previous steps, however, the relevant computation has already been performed in \cite{miles1984internally}.
The key observation is that at $\mathcal{O}(\epsilon^3)$ in the equations of motion, the new source terms only involve the eigenfunctions and eigenvalues corresponding to the vortex-free state.
Corrections due to $v_\theta$ will be suppressed by an additional power of $\epsilon$ and therefore go beyond our working order in perturbation theory.
Hence, we can use the nonlinear frequency correction in \cite{miles1984internally} for a propagating $m=1$ wave inside a cylindrical vessel,
\begin{equation}
\omega_\mathrm{nl} = -\frac{\chi k_1^2 \overline{\eta^2}}{2 \tanh^2 (h_0 k_1)},
\end{equation}
where $\chi$ depends only on the aspect ratio of the vessel $h_0/r_B$ ($\chi\simeq -0.4$ for our experiment which has $h_0/r_B\simeq 0.579$) and the overbar indicates the average over all space and (fast) time.
In this expression, we use the predicted profile of the primary mode, which agrees well with the data inside the field of view and extends over the full spatial domain, allowing us to easily evaluate the spatial integral in $\overline{\eta^2}$.
The nonlinear frequency correction shown to be consistent with the data in Fig. 2(a,b) of the main text.


\end{document}